\documentclass[10pt,oneside,conference,compsocconf]{amsart}
\usepackage[latin9]{inputenc}
\usepackage{amsthm}
\usepackage{graphicx}

\makeatletter

\providecommand{\tabularnewline}{\\}

\numberwithin{equation}{section}
\numberwithin{figure}{section}

\usepackage{mathtools}
\usepackage{algorithm}
\usepackage{algpseudocode}
\usepackage{color}% Set the graphics path for all images that will be imported.
\usepackage{epstopdf}

\usepackage{acronym}

\newacro{NSF}{National Science Foundation}
\newacro{VM}{Virtual Machine}
\newacro{AWS}{Amazon Web Services}
\newacro{EC2}{Amazon Elastic Compute Cloud}
\newacro{EMR}{Amazon Elastic MapReduce}
\newacro{ECU}{EC2 Compute Unit}
\newacro{S3}{Amazon Simple Storage Service}
\newacro{CLI}{Command Line Interface}
\newacro{SQL}{Structured Query Language}
\newacro{AP}{Affinity Propagation}
\newacro{HAP}{Hierarchical Affinity Propagation}
\newacro{MR-HAP}{MapReduce Hierarchical Affinity Propagation}
\newacro{HDFS}{Hadoop Distributed Filesystem}
\newacro{JVM}{Java Virtual Machine}
\newacro{ML}{Machine Learning}
\newacro{FOSS}{Free and Open Source Software}
\newacro{DBSCAN}{Density-Based Spatial Clustering of Applications with Noise}
\newacro{PRBP}{Partition with Reduce Boundary Points}
\newacro{HVKM}{Hierarchical Virtual K-means}
\newacro{D3.js}{Data-Driven Documents}
\newacro{HK-Means}{Hierarchical K-Means}

\newcommand{\ie}{\textit{i.e.}}
\newcommand{\eg}{\textit{e.g.}}
\newcommand{\etal}{et al.}

\makeatother

\begin{document}
\title{Parallel Hierarchical Affinity Propagation with MapReduce}
\author{Dillon Mark Rose}
\author{Jean Michel Rouly}
\author{Rana Haber}
\author{Nenad Mijatovic }
\author{Adrian M. Peter}

\begin{abstract}
The accelerated evolution and explosion of the Internet and social
media is generating voluminous quantities of data (on zettabyte scales).
Paramount amongst the desires to manipulate and extract actionable
intelligence from vast big data volumes is the need for scalable,
performance-conscious analytics algorithms. To directly address this
need, we propose a novel MapReduce implementation of the exemplar-based
clustering algorithm known as Affinity Propagation. Our parallelization
strategy extends to the multilevel Hierarchical Affinity Propagation
algorithm and enables tiered aggregation of unstructured data with
minimal free parameters, in principle requiring only a similarity
measure between data points. We detail the linear run-time complexity
of our approach, overcoming the limiting quadratic complexity of the
original algorithm. Experimental validation of our clustering methodology
on a variety of synthetic and real data sets (\eg\ images and point
data) demonstrates our competitiveness against other state-of-the-art
MapReduce clustering techniques. 
\end{abstract}

\keywords{MapReduce, Cluster, Affinity Propagation, Hierarchical Affinity Propagation,
Hadoop }

\maketitle

\section{Introduction}

In 2010, big data was growing at 2.5 quintillion~\cite{Humbetov12}
bytes per day. This overwhelming volume, velocity, and variety of
data can be attributed to the ubiquitously spread sensors, perpetual
streams of user-generated content on the web, and increased usage
of social media platforms---Twitter alone produces 12 terabytes of
tweets every day. The sustained financial health of the world's leading
corporations is intimately tied to their ability to sift, correlate,
and ascertain actionable intelligence from big data in a timely manner.
These immense computational requirements have created a heavy demand
for advanced analytics methodologies which leverage the latest in
distributed, fault-tolerant parallel computing architectures. Among
a variety of choices, MapReduce has emerged as one of the leading
parallelization strategies, with its adoption rapidly increasing due
to the availability of robust open source distributions such as Apache
Hadoop~\cite{hadoop13}. In the present work, we develop a novel
MapReduce implementation of a fairly recent clustering approach \cite{Givoni12},
and demonstrate its favorable performance for big data analytics.%
\footnote{Implementation available at http://research2.fit.edu/ice/?q=software%
} 

Clustering techniques are at the heart of many analytics solutions.
They provide an unsupervised solution to aggregate similar data patterns,
which is key to discovering meaningful insights and latent trends.
This becomes even more necessary, but exponentially more difficult,
for the big data scenario. Many clustering solutions rely on user
input specifying the number of cluster centers (e.g. K-Means clustering~\cite{hartigan78}
or Gaussian Mixture Models~\cite{McLachlan00}), and biasedly group
the data into these desired number of categories. Frey \etal\ \cite{Frey07}
introduced an exemplar-based clustering approach called \ac{AP}.
As an exemplar-based clustering approach, the technique does not seek
to find a mean for each cluster center, instead certain representative
data points are selected as the exemplars of the clustered subgroups.
The technique is built on a message passing framework where data points
``talk'' to each other to determine the most likely exemplar and
automatically determine the clusters, \ie\ there is no need to specify
the number of clusters a priori. The sole input is the pairwise similarities
between all data points under consideration for clustering---making
it ideally suited for a variety of data types (categorical, numerical,
textual, etc.). A recent extension of the \ac{AP} clustering algorithm
is \acf{HAP}~\cite{Givoni12}, which groups and stacks data in
a tiered manner. \ac{HAP} only requires the number of hierarchy levels
as input and the communication between data points occurs both within
a single layer and up and down the hierarchy. To date, \ac{AP}
and \ac{HAP} have been mainly relegated to smaller, manageable
quantities of data due to the prohibitive quadratic run time complexity.
Our investigations will demonstrate an effective parallelization strategy
for \ac{HAP} using the MapReduce framework, for the first time
enabling applications of these powerful techniques on big data problems.

First introduced by Google~\cite{dean04}, the MapReduce framework
is a programming paradigm designed to facilitate the distribution
of computations on a cluster of computers. The ability to distribute
processes in a manner that takes the computations to the data is key
when mitigating the computational cost of working with extremely large
data sets. The parallel programming model depends on a mapper phase
that uses key-value identifiers for the distribution of data and subsequent
independent executions to generate intermediate results. These are
then gathered by a reducing phase to produce the final output key-value
pairing. This simple, yet widely applicable parallelization philosophy
has empowered many to take machine learning algorithms previously
demonstrated only on ``toy data'' and scale them to enterprise-level
processing~\cite{Ng07}. In this same vein, we adopted the most popular
open source implementation of the MapReduce programming model, Apache
Hadoop, to develop the \emph{first ever} parallelized extension of
\ac{HAP}, which we refer to as \ac{MR-HAP}. This allows efficient
fault-tolerant clustering of big data, and more importantly, improves
the run time complexity to potentially linear time (given enough machines).

\subsection{Relevant Work}

To handle the explosion of available data, there is now a vast amount
of research in computational frameworks to efficiently manage and
analyze these massive information quantities. Here we focus on the
MapReduce framework~\cite{Humbetov12,Zewen12,Wang12,Esteves12} for
faster and more efficient data mining, covering the most relevant
to our approach.

Among the state of the art MapReduce clustering algorithms is \ac{HVKM},
which was implemented by Nair \etal\ \cite{Nair11}. \ac{HVKM}
uses cloud computing to handle large data sets, while supporting top
to bottom hierarchies or a bottom to top approach. Since it derives
its roots from K-means, \ac{HVKM} requires one to specify the number
of clusters. Our \ac{MR-HAP} implementation does not require presetting
the number of required clusters; it instead organically and objectively
discovers the data partitions.

In Wu \etal\ \cite{Wu10}, the authors propose to parallelize \ac{AP}
on the MapReduce framework to cluster large scale E-learning resources.
The parallelization happens on the individual message level of the
\ac{AP} algorithm. We perform a similar parallelization but significantly
go beyond and allow for hierarchical clustering, which enables a deeper
understanding of the data's semantic relationships. In addition, our
development is designed to work on a variety of data sources; thus,
our experiments will showcase results on multiple data modalities,
including images and numerical data, as shown in \S~\ref{sec:Experimental-Results}.

The rest of this paper is organized as follows. In the next section,
\S\ref{sec:HAP}, we detail the non-parallel \ac{HAP} algorithm.
\S\ref{sec:MapReduce} discusses the MapReduce paradigm and the implementation
details for these algorithms. The experimental validations provided
in \S\ref{sec:Experimental-Results} demonstrate our favorable performance
against another clustering algorithm which is readily available in
the open source project Apache Mahout~\cite{mahout13}. Finally,
we conclude with a summary of our efforts and future recommendations.

\section{Hierarchical Affinity Propagation\label{sec:HAP}}

\ac{AP} is a clustering algorithm introduced by Frey \etal \cite{Frey07}
motivated by the simple fact that given pairwise similarities between
all input data, one would like to partition the set to maximize the
similarity between every data point and its cluster's exemplar. Recall
that an exemplar is an actual data point that has been selected as
the cluster center. As we will briefly discuss, these ideas can be
represented as an algorithm in a message passing framework. In the
landscape of clustering methodologies, which includes such staples
as K-means \cite{hartigan78}, K-medoids \cite{kaufman87}, and Gaussian
Mixture Models \cite{McLachlan00}, predominantly all methods require
the user to input the desired number of cluster centers. \ac{AP}
avoids this artificial segmentation by allowing the data points to
communicate amongst themselves and organically give rise to a partitioning
of the data. In many applications an exemplar-based clustering technique
gives each cluster a more representative and meaningful prototype
for the center, versus a fabricated mean.

\ac{HAP}, introduced by \cite{Givoni12}, extends \ac{AP} to allow
tiered clustering of the data. The algorithm starts by assuming that
all data points are potential exemplars. Each data point is viewed
as a node in a network connected to other nodes by arcs such that
the weight of the arcs $s_{ij}$ describes how similar the data point
with index $i$ is to the data point with index $j$. \ac{HAP} takes
as input this similarity matrix where the entries are the negative
real valued weights of the arcs. Having the similarity matrix as the
main input versus the data patterns themselves provides an additional
layer of abstraction---one that allows seamless application of the
same clustering algorithm regardless of the data modality (\eg  text,
images, general features, etc.). The similarity can be designed to
be a true metric on the feature space of interest or a more general
non-metric \cite{Frey07}. The negative of the squared Euclidean distance
if often used as a metric for the similarities. The diagonal values
of the similarity matrix, $s_{jj}$, are referred to as the \textquotedblleft{}preferences\textquotedblright{}
which specify how much a data point j wants to be an exemplar. Since
the similarity matrix entries are all negative values, $-\infty<s_{ij}\le0$,
$s_{jj}=0$ implies data point $j$ has high preference of being an
exemplar and $s_{jj}\approx-\infty$ implies it has very low preference.
In some cases, as in \cite{Frey07,Givoni12,Xiao07}, the preference
values are set using some prior knowledge; for example, uniformly
setting them to the average of the maximum and minimum values of $s_{ij}$,
or by setting them to random negative constants. Through empirical
verification, we experienced better performance with randomizing the
preferences and adopt this approach for most of our experiments. Once
the similarity matrix is provided to the \ac{HAP} algorithm, the
network of nodes (data points) recursively transmits two kinds of
intra-level messages between each node until a good set of exemplars
is chosen. The first message is known as the \textquotedblleft{}responsibility\textquotedblright{}
message and the second as the \textquotedblleft{}availability\textquotedblright{}
message. The responsibility messages, $\rho_{ij}^{l}$, are sent at
level $l$ from data point $i$ to data point $j$ portraying how
suitable node $i$ thinks node $j$ is to be its exemplar. Similarly,
availability messages, $\alpha_{ij}^{l}$, are sent at level $l$
from data point $j$ to $i$, indicating how available $j$ is to
be an exemplar for data point $i$. The responsibility and availability
update equations are given in Eq. \ref{eq:RespUpdateHAP} and Eq.
\ref{eq:AvailUpdateHAP}, respectively. 
\begin{eqnarray}
\rho_{ij}^{l>1} & \leftarrow & s_{ij}^{l}+\min[\tau_{i}^{l},-\max_{ks.t.k\neq j}\left\{ \alpha_{ik}^{l}+s_{ik}^{l}\right\} ]\label{eq:RespUpdateHAP}\\
\alpha_{ij}^{l<L} & \leftarrow & \min\Big\{0,c_{j}^{l}+\phi_{j}^{l}+\rho_{jj}^{l}+\sum_{\mathclap{ks.t.k\notin\{i,j\}}}\max\{0,\rho_{kj}^{l}\}\Big\}\label{eq:AvailUpdateHAP}\\
\alpha_{jj}^{l<L} & \leftarrow & c_{j}^{l}+\phi_{j}^{l}+\sum_{ks.t.k\neq j}\max\{0,\rho_{kj}^{l}\}\label{eq:self-availabilityHAP}
\end{eqnarray}
where $L$ is the number of levels defined by the user and $l\in\{1,\cdots,L\}$.
Eq. \ref{eq:self-availabilityHAP} is the self-availability equation
which reflects the accumulated positive evidence that $j$ can be
an exemplar. The self-responsibility messages are updated the same
way as the responsibility messages. To avoid numerical oscillation,
the responsibility and availability messages are dampened by $\lambda\in(0,1)$
at every level $l$.

\ac{HAP} also introduces two inter-level messages. These messages
are denoted by $\tau$ in Eq. \ref{eq:tau}, which receives messages
from the lower level and $\phi$ in Eq. \ref{eq:Phi}, which receives
messages from the upper level. At every level, the cluster preference
$c_{i}^{l}$ is updated using Eq. \ref{eq:exemplarLevelUpdate}.
\begin{eqnarray}
\tau_{j}^{l+1} & = & c_{j}^{l}+\rho_{jj}^{l}+\sum_{ks.t.k\neq j}\max(0,\rho_{kj}^{l})\label{eq:tau}\\
\phi_{i}^{l-1} & = & \max_{k}(\alpha_{ik}^{l}+s_{ik}^{l})\label{eq:Phi}\\
c_{i}^{l} & \leftarrow & \max_{j}(\alpha_{ij}^{l}+\rho_{ij}^{l})\label{eq:exemplarLevelUpdate}
\end{eqnarray}

A variety of strategies can be employed to update the similarity matrix
$s_{ij}^{l}$ to vary level-wise. We have achieved good results by
simply taking into consideration the cluster relationship of the previous
level:
\begin{equation}
s_{ij}^{l+1}=s_{ij}^{l}+\kappa\max_{js.t.j\neq i}[\alpha_{ij}^{l}+\rho_{ij}^{l}]\label{eq:SimilarityLevelUpdate}
\end{equation}
where $\kappa$ is a constant value within {[}0,1{]}. This updates
the relation between data points in level $l+1$ by negatively increasing
the similarity between points that belong to different clusters in
level $l$ and enforces the similarity between points that fall under
the same cluster in level $l$.

After all messages have been sent and received, the cluster assignments
are chosen, at every level, based on the maximum sum of the availability
and responsibility messages as in Eq. \ref{eq:exemplarPositionLevelUpdate}.
These cluster assignments can be used to extract the list of exemplars.
\begin{eqnarray}
e_{i}^{l} & \leftarrow & \arg\max_{j}\{\alpha_{ij}^{l}+\rho_{ij}^{l}\}\label{eq:exemplarPositionLevelUpdate}
\end{eqnarray}
 These net message exchanges seek to maximize the cost of correctly
labeling a point as an exemplar and gathering its representative members
(a cluster). In Algorithm 1, we detail the pseudo-code implementation
of \ac{HAP}. Given this description of \ac{HAP}, we now proceed
to discuss MapReduce and our novel parallelization strategy.

\begin{algorithm}[t]
\caption{Hierarchical Affinity Propagation} 
\begin{algorithmic}[1] 
\State \textbf{Input:} Similarity (S), Levels (L), Iterations, and $\lambda$ 
\State \textbf{Initialize:} $\alpha=0$, $\rho=0$, $\tau=\infty$, $\phi=0$, $c=0$, $e=0$
\For {$iter=1 \to$ Iterations} 
\For {$l=1 \to$ Levels}  		
\State Update $\rho^l_{ij}$ (eq.~\ref{eq:RespUpdateHAP}) \& Dampen $\rho^l$
\State Update $\alpha^l_{ij}$ (eq.~\ref{eq:AvailUpdateHAP} \& \ref{eq:self-availabilityHAP}) \& Dampen $\alpha^l$
\State Update $\tau^l_j$, $\phi^l_j$ \& $c^l_j$ (eq.~\ref{eq:tau}, \ref{eq:Phi} \& \ref{eq:exemplarLevelUpdate})			
\State \textbf{Optional} Update $s^l_{ij}$ (eq.~\ref{eq:SimilarityLevelUpdate}) 
\EndFor 
\EndFor 
\For {$l=1 \to$ Levels}
\State Update $e^l_j$ (\ref{eq:exemplarPositionLevelUpdate})
\EndFor
\label{alg:HAP}
\end{algorithmic} 
\end{algorithm}

\section{MapReduce Hierarchical Affinity Propagation\label{sec:MapReduce}}

The MapReduce programming model~\cite{dean04} is an abstract programming
paradigm independent of any language that allows the processing workload
of the implemented algorithm to be balanced over separate nodes within
a computer cluster. Our overarching MapReduce approach for \ac{HAP}
was motivated by viewing the major update equations for \ac{HAP}
(see Algorithm 1) as tensorial mathematical constructs~\cite{Lu11}.
One can simply view these tensorial constructs as two or three dimensional
matrices. The \ac{HAP} algorithm can be parallelized because all
the updates to the various tensors require only a subset of the information
provided. Therefore, the updates can be split up into different jobs
and each job will receive the subset of data needed to evaluate the
update.

To achieve a balance between computational partitioning and efficient
formatting for data representation on the \ac{HDFS}, all the data
is constructed as three dimensional tensors. In support of the fault
tolerance aspect of MapReduce, it is important to retain a copy at
all times of the $S$, $\alpha$, $\rho$, $c$, $\tau$, and $\phi$
tensors. (Recall $S$, $\alpha$, $\rho$, and $c$ refer to the Similarity,
Availability, Responsibility, and Cluster Preferences, respectively.)
To this end, even those tensors not required by a job must be passed
directly through to the next job. For the $S$, $\alpha$, and $\rho$
tensors, the dimensions represent the nodes, the exemplars, and the
levels. Since there are $N$ nodes, $N$ possible exemplars, and $L$
levels, these tensors contain $LN^{2}$ values. For the $c$, $\tau$,
and $\phi$ tensors, the first two dimensions represent the index
and level and the depth dimension has length one. Since there are
$N$ indices and $L$ levels, these tensors contain $LN$ values.
In the sequel, for the $S$, $\alpha$, and $\rho$ tensors, the node
dimension will be iterated by $i$, the exemplar dimension will be
iterated by $j$, and the level dimension iterated by $l$. As for
the $c$, $\tau$, and $\phi$ tensors, the index dimensions will
be iterated by both $i$ and $j$.

With these underlying structures, data must be deconstructed and represented
as (key,value) pairs for use in the MapReduce framework. There are
two formats for storing the information: node-based and exemplar-based
formatting. In the \emph{node-based format}, the keys are string tuples,
$(i,l,\xi)$, where $i$ represents the node, $l$ represents the
level, and $\xi$ represents the tensor $(\alpha,\rho,...)$. The
values, represented by $\nu$, are the vectors for the $i^{th}$ node
of the matrix on the $l^{th}$ level of the tensor. In the \emph{exemplar-based
format}, the keys are string tuples, $(j,l,\xi)$, where $j$ represents
the exemplar, $l$ represents the level, and $\xi$ represents the
tensor. The values, represented by $\nu$, are the vectors for the
$j^{th}$ exemplar of the matrix on the $l^{th}$ level of the tensor.
With the data thus represented, MapReduce jobs must be constructed
to manipulate the information using the given \ac{HAP} equations.

\begin{figure}
\begin{centering}
\includegraphics[width=0.3\textwidth]{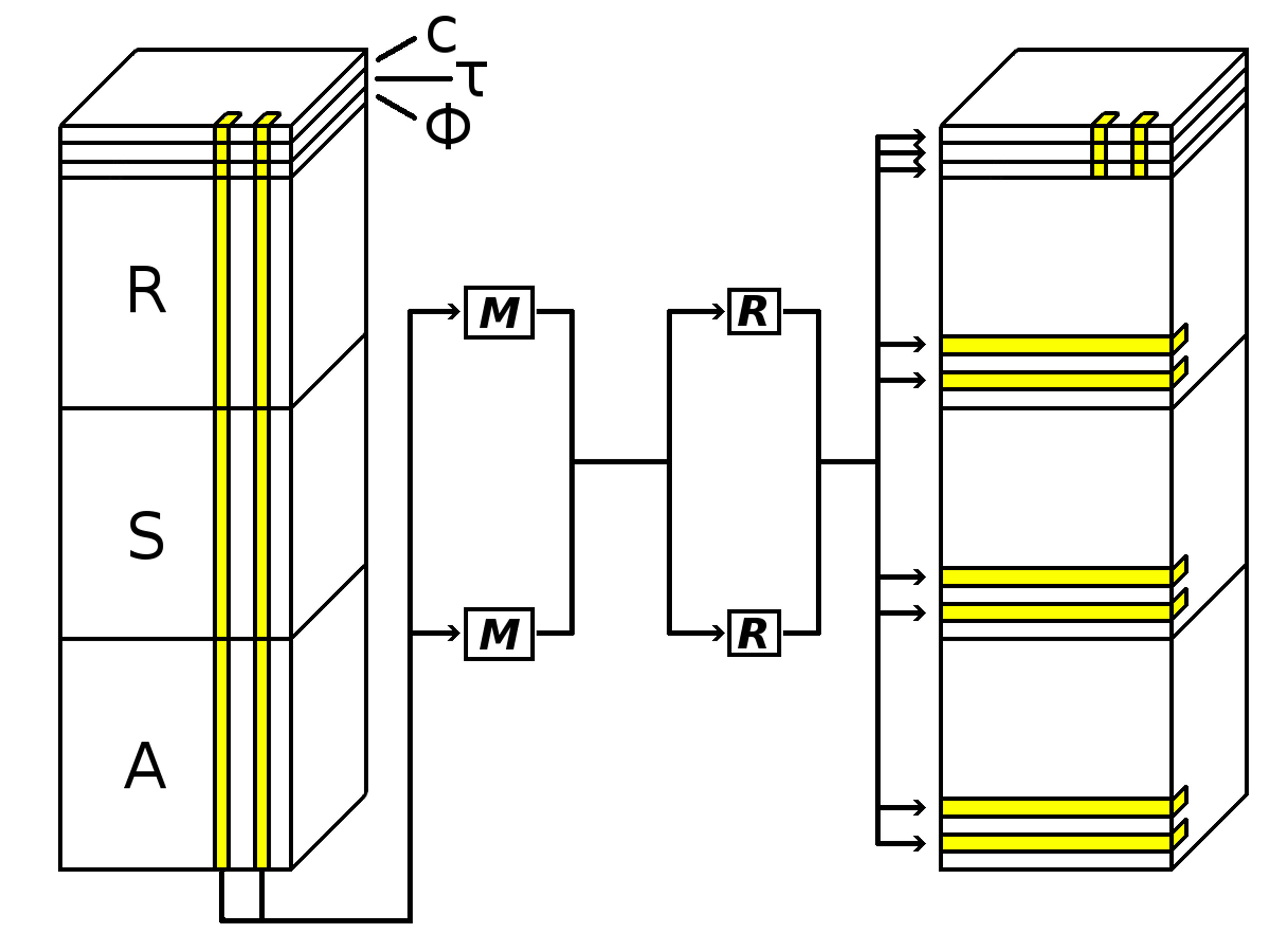} 
\par\end{centering}

\caption{\label{fig:Parallelization-Scheme}Parallelization Scheme}
\vspace{-10pt}
\end{figure}

In our parallelization scheme, \ac{MR-HAP} is broken down into
three separate MapReduce jobs. The first job handles updating $\tau$,
$c$, and $\rho$. The second job handles updating $\phi$ and $\alpha$.
These first two jobs loop for a set number of iterations. At the end
of the iterations, the final job extracts the cluster assignments
on each level. Due to dependencies set out in the equations, the $\rho$
update must occur first. Therefore, $\tau$ and $c$ are not updated
during the first iteration. In all other iterations they occur before
the Responsibility update. At the start of each iteration, the data
will be in exemplar-based format. After the first job, the data will
have switched to node-based format. The second job converts the data
back to exemplar-based format to begin a new iteration or to be used
as input to the final job. See Fig.~\ref{fig:Parallelization-Scheme}
for a visual representation of the parallelization scheme. The figure
represents what happens to the data during either of the first two
jobs. The tensors have been stacked to show how the indices line up.
The yellow strips on the left represent information being passed to
one mapper, one strip per mapper. The focus of each mapper is on providing
the reducers with the necessary information. The subsequent focus
of each reducer is on performing the tensor updates as defined in
the \ac{HAP} equations. As the data comes out of the job, the switch
between exemplar-based and node-based formats can be easily seen.
The output is now ready for use by the next job, which will follow
a similar flow. The following sections will provide in-depth explanations
of each MapReduce job.

\subsubsection{Updating $\tau$, $c$, and $\rho$}

This job takes as input the exemplar-based representation of the data
and outputs the node-based representation of the data with updated
values. In the first iteration, $\tau$ and $c$ are not updated due
to previously mentioned dependencies. In this MapReduce job, the mapper
deconstructs the exemplar-based vectors into node-based values for
the reducer to reconstruct node-based vectors. Each mapper receives
a key describing a unique $(j,l,\xi)$ combination and a value with
the corresponding vector. The indices of the vector represent the
nodes; thus, the mapper iterates over the vector with $i$. Each reducer
receives a key describing a unique $(i,l)$ combination and a list
of values which will be used to reconstruct the 6 node-based vectors,
the 2 node-based vectors from the level below and the 2 special diagonal
vectors. The indices of the constructed vector represent the exemplars
so the reducer iterates over the vector with $j$.

\subsubsection{Updating $\alpha$ and $\phi$}

This job takes as input the node-based representation of the data
and outputs the exemplar-based representation of the data with updated
values. In this MapReduce job, the mapper deconstructs the node-based
vectors into exemplar-based values for the reducer to reconstruct
exemplar-based vectors. Each mapper receives a key describing a unique
$(i,l,\xi)$ combination and a value with the corresponding vector.
The mapper iterates over the vector with $j$. Each reducer receives
a key describing a unique $(j,l)$ combination and a list of values
which will be used to reconstruct the 6 exemplar-based vectors and
the 2 node-based vectors from the level above. The indices of the
constructed vector represent the nodes so the reducer iterates over
the vector with $i$.

\subsubsection{Extracting Cluster Assignments}

This job takes as input the exemplar-based representation of the data
and outputs the cluster assignments. In this MapReduce job, the mapper
deconstructs the exemplar-based vectors into node-based values for
the reducer to reconstruct node-based vectors. Each mapper receives
a key describing a unique $(j,l,\xi)$ combination and a value with
the corresponding vector. The mapper iterates over the vector with
$i$. Since this is the last step, only the required information has
to pass to the reducer and the other information can be neglected.
Each reducer receives a key describing a unique $(i,l)$ combination
and a list of values which will be used to reconstruct the 2 node-based
vectors and the 2 special diagonal vectors. The reducer iterates over
the vector with $j$.

\subsection{Runtime Complexity\label{sec:RuntimeComplexity}}

A standard sequential \ac{HAP} implementation must necessarily
have a runtime complexity of $O(kLN^{2})$ where $k$ represents the
number of algorithmic iterations, run either as a hard limit or until
convergence is reached, $L$ represents the number of output levels
requested, and $N$ represents the cardinality of the input data set,
such that the size of $S$ is $(L\times N\times N)$. The runtime
complexity is a direct result of iterating over all three dimensions
of the tensors for each iteration. By implementing the algorithms
in the MapReduce framework, we are able to achieve superior runtime
complexity. Under MapReduce, the \ac{MR-HAP} runtime complexity
reduces to a linear relationship with the data, assuming the total
number of \acp{VM} on the cluster, $M$, scales to $LN$, \ie\ $O(\frac{kLN^{2}}{M})=O(kN)$
as $M\to LN$. In \ac{MR-HAP}, $M$ can only scale up to a maximum
of $LN$ because $M$ is limited to the number of tasks that can be
evaluated at the same time. In this case, it is limited to the minimum
of the $6LN$ mapper tasks and the $LN$ reducer tasks, where the
constant factor six represents the number of tensor identifications
introduced into the algorithm, namely $\alpha,\rho,S,\tau,\phi$,
and $c$.

\section{Experimental Results\label{sec:Experimental-Results}}

\begin{figure}
\begin{centering}
\begin{tabular}{cc}
\includegraphics[scale=0.11325]{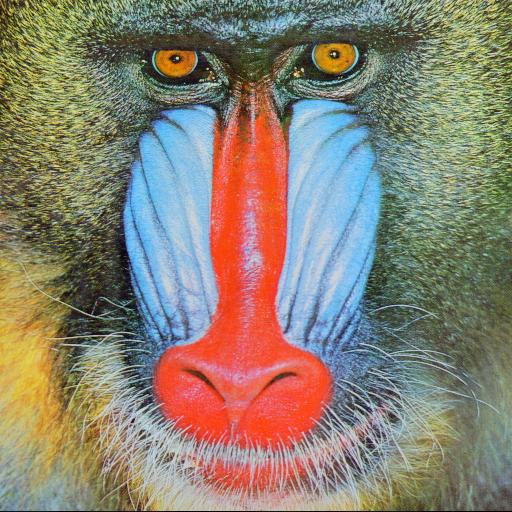}  & \includegraphics[scale=0.75]{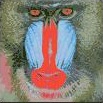}\tabularnewline
a) Original ``Mandrill'' \label{fig: Baboon-a}  & b) 15 Exemplars\label{fig: Baboon-b}\tabularnewline
\includegraphics[scale=0.75]{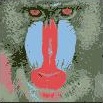}  & \includegraphics[scale=0.75]{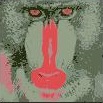}\tabularnewline
c) 7 Exemplars\label{fig: Baboon-c}  & d) 6 Exemplars\label{fig: Baboon-d}\tabularnewline
\end{tabular}
\par\end{centering}

\caption{\label{fig:Baboon}Hierarchical clustering of ``Mandrill'' 103x103.
See text for discussion.}
\vspace{-10pt}
\end{figure}
To demonstrate the effectiveness and adaptability of the proposed
approach, we executed validation experiments on several data sets
with a variety of modalities, \eg\ imagery and synthesized numerical
point data. Where applicable, we compared our performance to a popular
MapReduce hierarchical clustering algorithm currently available in
the Mahout library. At its core, their hierarchical clustering is
based on a level-wise K-means clustering approach; thus, we refer
to it as \ac{HK-Means}. With K-means as the foundation, \ac{HK-Means}
requires the number of cluster centers as input. Since our method
does not explicitly impose this requirement, we adopted the initialization
method of running Canopy clustering, also available in Mahout, to
discover the ``natural'' number of centers. We then use these cluster
centers to seed \ac{HK-Means}. In order to truly gain an objective
understanding of \ac{MR-HAP} performance versus \ac{HK-Means},
we use the \emph{purity extrinsic cluster quality metric} to assess
their respective aggregation capabilities \cite{Sahoo06}.

\subsection{Image Segmentation}

\begin{figure}
\begin{centering}
\begin{tabular}{cc}
\includegraphics[scale=0.6]{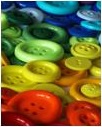}  & \includegraphics[scale=0.6]{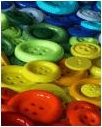}\tabularnewline
a) Original ``Buttons'' \label{fig: Buttons-a}  & b) 154 Exemplars\label{fig: Buttons-b}\tabularnewline
\includegraphics[scale=0.6]{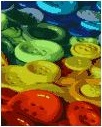}  & \includegraphics[scale=0.6]{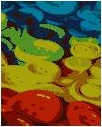}\tabularnewline
c) 25 Exemplars\label{fig: Buttons-c}  & d) 11 Exemplars\label{fig: Buttons-d}\tabularnewline
\end{tabular}
\par\end{centering}

\caption{\label{fig:Buttons}Hierarchical clustering of ``Buttons'' 120x100.
See text for discussion.}
\vspace{-10pt}
\end{figure}

Hierarchical Affinity Propagation performs very well in image segmentation
tasks as shown in Fig.~\ref{fig:Baboon} \& Fig.~\ref{fig:Buttons}.
The ``Mandrill'' image, Fig.~\ref{fig:Baboon}, is of size $103\times103$,
which provided 10,609 pixels (data points) to cluster. Similarly,
the ``Buttons'' image, Fig.~\ref{fig:Buttons}, is of size $120\times100$,
resulting in a data set of 12,000 pixels. The similarity input was
computed using the negative Euclidean distance between all pixels
treating RGB intensities as vectors. The diagonal, or preference entries,
were selected as random numbers within $[-10^{6},0]$. As for the
other parameters, we set the iterations to 30 and the dampening factor
to $\lambda=0.5$. To generate the clustered images, we re-color all
pixels within a cluster with the color of the selected exemplar. The
number of hierarchy levels for the ``Mandrill'' data set was set
to $L=3$. The top right image is the lowest level where the pixels
were grouped into 15 clusters. The bottom left image is the second
level where the pixels were grouped into 7 clusters. Finally, the
bottom right image is the highest level where the pixels were grouped
into 6 clusters. From these images we can still see the mandrill's
shape and most of its colors, but at the highest level it appears
fuzzier. This is because the members of the same clusters were given
the color of the exemplar.

For the ``Buttons'' image, the number of levels was set to $L=3$.
The top right image is the lowest level where the pixels were grouped
into 154 clusters. The bottom left image is the second level where
the pixels were grouped into 25 clusters. Finally, the bottom right
image is the highest level where the pixels were grouped into 11 clusters.
The highest level of the hierarchy appears fuzzier than the original
image due to similar colors clustering underneath a single exemplar.

\subsection{Scalability and Comparison to HK-Means \label{sec:Performance}}

In order to test the scalability of the \ac{MR-HAP} algorithm with
respect to speed, we use the data set ``Aggregation''~\cite{Gionis07},
which is a shape set composed of 788 two-dimensional points. The purpose
of these tests was to observe trends in algorithm runtime as cluster
computing power increased, as well as to determine the benefits of
running in a distributed environment as compared to an undistributed
environment (a single-machine Hadoop cluster). Hadoop clusters were
provided using \ac{EMR} to dynamically create clusters of standard
\ac{EC2} instances. Cluster computing power was scaled both by
increasing the number of \acp{VM} within a cluster and by provisioning
more powerful \acp{VM}. The two \ac{VM} instance types used
are: (1) the m1.small, which has 1.7 GB of memory and is considered
to have 1 \ac{ECU} with 160 GB of instance storage and a 32-bit
architecture, and (2) the m1.xlarge, which has 15 GB of memory, 8
\ac{ECU}, 1,690 GB of instance storage, and a 64-bit architecture.
The single-machine Hadoop cluster utilized to simulate an undistributed
environment has 8 GB of memory, 8 \ac{ECU}, 40 GB of machine storage,
and a 64-bit architecture.

\begin{figure}
\centering{}\includegraphics[scale=0.2]{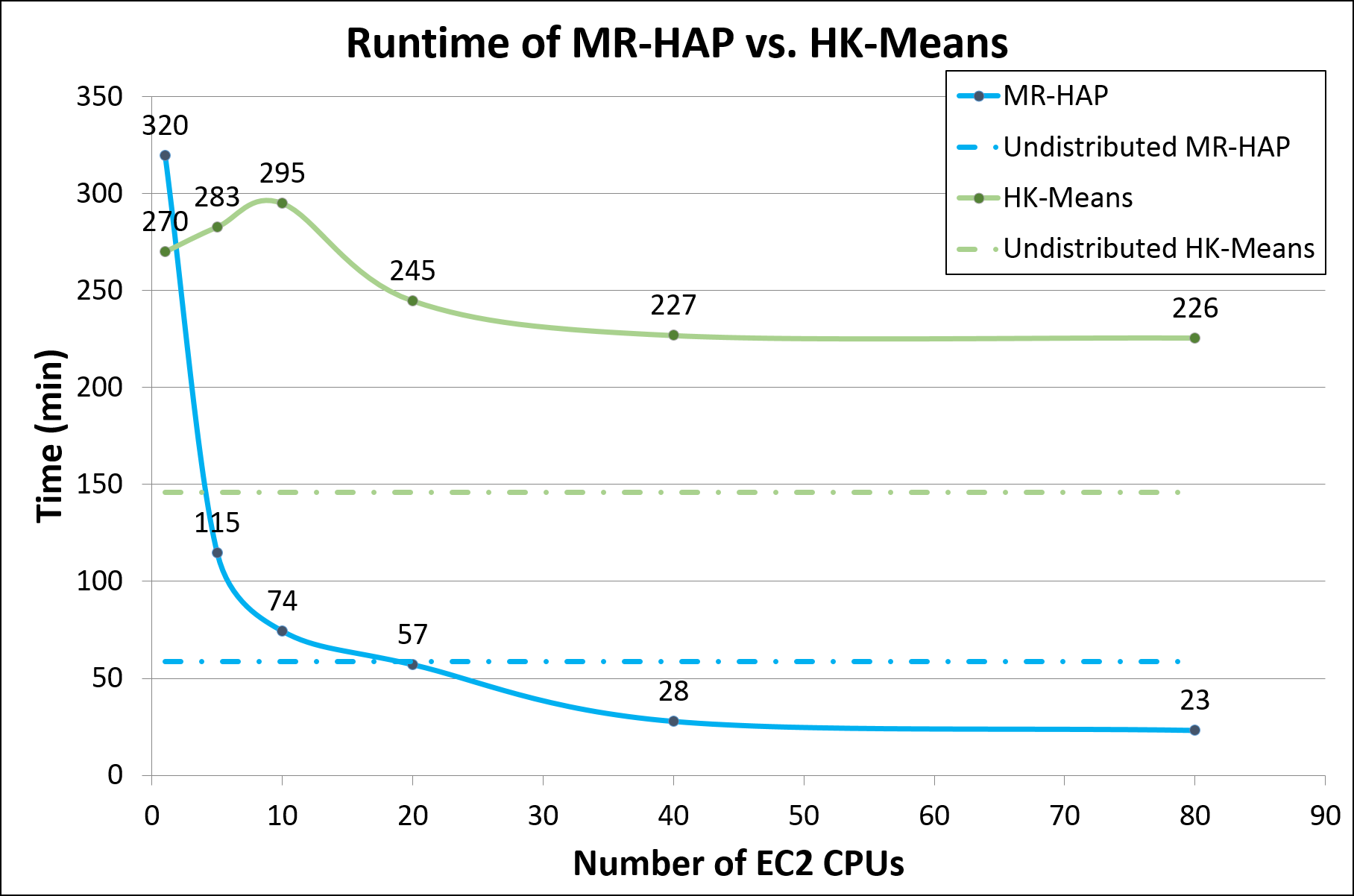}\caption{\label{fig:Speed}Time vs. Number of EC2 CPUs. Our \ac{MR-HAP}
better utilizes available compute resources to significantly improve
runtime.}
\vspace{-10pt}
\end{figure}

\begin{figure}
\centering{}\includegraphics[scale=0.2]{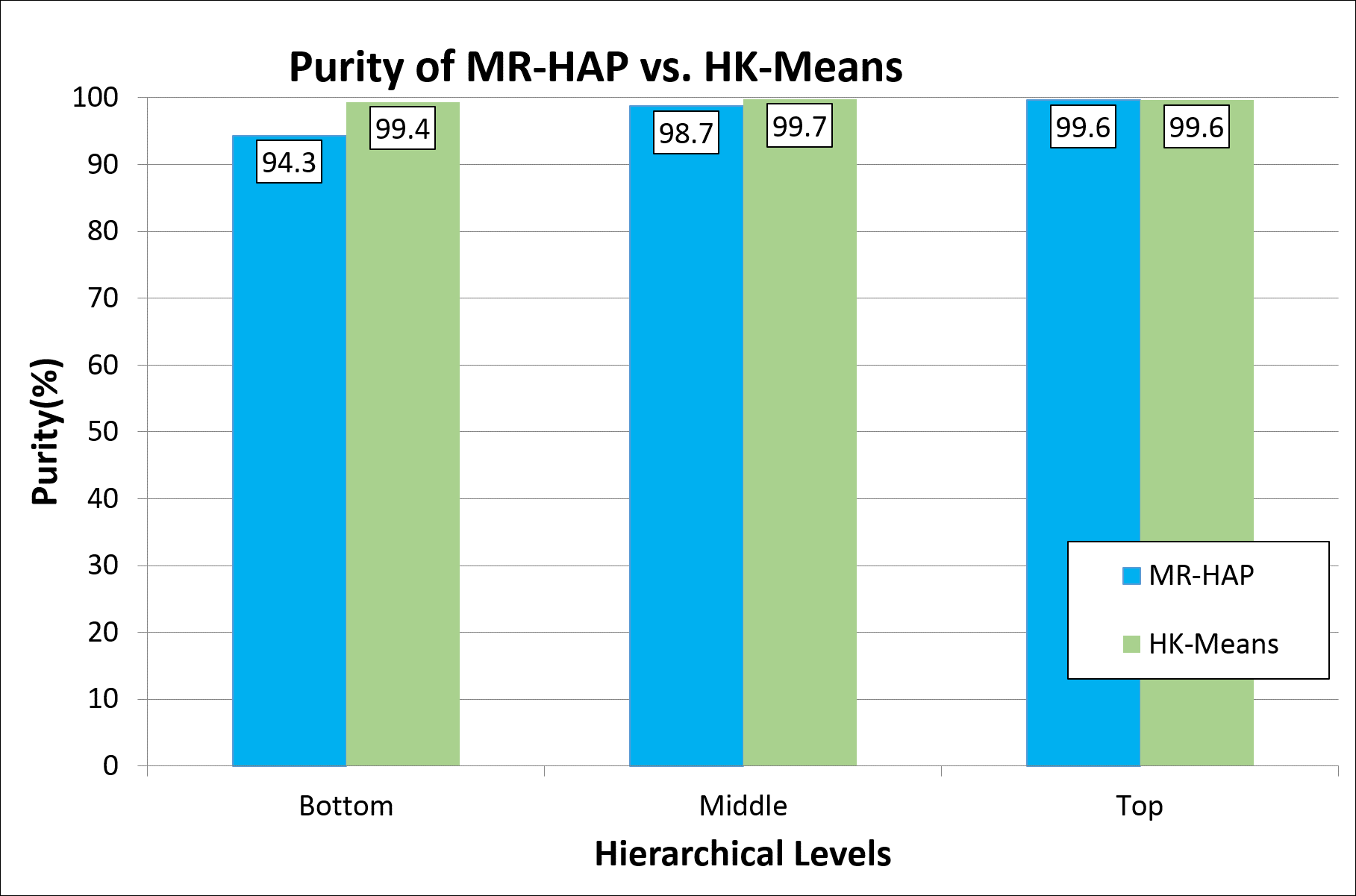}\caption{\label{fig:Purity}Purity levels of \ac{MR-HAP} vs. \ac{HK-Means}.
\ac{MR-HAP} posts results highly competitive with \ac{HK-Means}.}
\vspace{-10pt}
\end{figure}

For comparison to another state-of-the-art MapReduce clustering methodology,
our \ac{MR-HAP} algorithm was benchmarked against \ac{HK-Means}.
Due to its inherently parallel design, \ac{MR-HAP} immediately
begins to benefit from being placed in a distributed environment.
Represented by a solid blue line in Fig.~\ref{fig:Speed}, \ac{MR-HAP}
runtime decreases by 64\%, from 320 minutes to 115 minutes, when cluster
computational power is increased by just 4 additional \ac{ECU}.
\ac{MR-HAP} eventually reaches the threshold of a linear relationship
with the size of the input data at a runtime of around 20 minutes,
which is a 94\% decrease from the single \ac{ECU} cluster. Furthermore,
at its best, \ac{MR-HAP} performs 66\% faster in a distributed
environment than the undistributed environment which is represented
by the blue dotted line in Fig.~\ref{fig:Speed}. In contrast, the
Mahout HK-Means algorithm used in this experimentation, indicated
by the solid green line in Fig.~\ref{fig:Speed}, is not parallelized
to the extent of \ac{MR-HAP}. Each single iteration of K-Means
is structured under Mahout to distribute over a Hadoop cluster, but
the hierarchical ``Top Down'' structure requires iterative executions
of K-Means for each level. This lack of an overall parallelization
scheme results in reduced performance at scale than \ac{MR-HAP}.
HK-Means runtime initially increases by 8.5\% when \ac{ECU} is
increased from 1 to 10 due to Hadoop cluster overhead, including network
latency and I/O time. However, at 10 \ac{ECU}, \ac{HK-Means}
overcomes this overhead and begins to benefit from the MapReduce parallelization
scheme. This results in an eventual 16\% runtime decrease between
1 and 80 \ac{ECU}, at which point \ac{HK-Means} eventually reaches
a linear relationship with the data at a runtime around 225 minutes.
Unlike \ac{MR-HAP}, HK-Means never surpasses its undistributed
runtime threshold of 146 minutes indicated by the green dotted line
in Fig.~\ref{fig:Speed}. Finally, at its best, \ac{HK-Means}
runs 90\% slower than \ac{MR-HAP}, requiring 226 minutes of execution
compared to \ac{MR-HAP}'s 23 minutes. With significantly faster
runtimes, \ac{MR-HAP} still posts purity levels competitive with
HK-Means, shown in Fig.~\ref{fig:Purity}. This combination of speed
and high performance is ideal for processing big data in a large-scale
cloud computing environment.

\section{Conclusion}

The need for efficient and high performing data analysis frameworks
remain paramount to the big data community. The \ac{AP} clustering
algorithm is rapidly becoming a favorite amongst data scientists due
to its high quality grouping capabilities, while requiring minimal
user specified parameters. Recently, a multilayer structured version
of the \ac{AP} algorithm, \ac{HAP}, was introduced to automatically
extract tiered aggregations inherent in many data sets. \ac{HAP}
is modeled as a message-based network that allows communication between
nodes and between levels in the hierarchy, and mitigates many of the
biases that arise in techniques that require one to input the number
of clusters. In the present work, we have developed the first ever
extension, \ac{MR-HAP}, to address the big data problem---demonstrating
an efficient parallel implementation using MapReduce that directly
improves the runtime complexity from quadratic to linear. The novel
tensor-based partitioning scheme allows for parallel message updates
and utilizes a consistent data representation that is leveraged by
map and reduce tasks. Our approach seamlessly allows us to cluster
a variety of data modalities, which we experimentally showcased on
data sets ranging from synthetic numerical points to imagery. Our
analysis and computational performance is competitive with the state-of-the-art
in MapReduce clustering techniques.

\subsection*{Acknowledgment:}

The authors acknowledge partial support from NSF grant No. 1263011. Any opinions, findings, and conclusions or recommendations expressed in this material are those of the authors and do not necessarily reflect the views of the NSF.

\small{

\bibliographystyle{IEEEtran}
\bibliography{RefList}

}
\end{document}